\documentclass[submission]{eptcs}

\usepackage[font=footnotesize,caption=false]{subfig}
\usepackage{graphicx}
\usepackage{amsfonts}
\usepackage{mathrsfs}
\usepackage{rotating}
\usepackage{fixltx2e}
\usepackage{float}
\usepackage{xspace}
\usepackage{soul}
\usepackage{caption}
\usepackage{cite}

\providecommand{\event}{TAV-WEB 2010} 

\newcommand{\rumor}{\textsc{RuMoR}\xspace}
\newcommand{\tbs}{$\mathtt{TBS}$\xspace}

\newtheorem{mydefinition}{Definition}
\newtheorem{myobservation}{Observation}
\newtheorem{myproposition}{Proposition}
\newtheorem{myclaim}{Claim}
\newtheorem{mylemma}{Lemma}
\newtheorem{mycorollary}{Corollary}
\newtheorem{myexample}{Example}
\newtheorem{myalgorithm}{Algorithm}

\newcommand{\bolddot}{\hspace{-1.5mm}\textbf{.}\ \  }

\newcommand{\BT}{\begin{theorem}}
\newcommand{\ET}{\end{theorem}}
\newcommand{\BCR}{\begin{mycorollary}\bolddot}
\newcommand{\ECR}{\end{mycorollary}}
\newcommand{\BPR}{\begin{myproposition}\bolddot}
\newcommand{\EPR}{\end{myproposition}}
\newcommand{\BL}{\begin{mylemma}\bolddot}
\newcommand{\EL}{\end{mylemma}}
\newcommand{\BCM}{\begin{myclaim}\bolddot}
\newcommand{\ECM}{\end{myclaim}}

\newcommand{\BD}{\begin{mydefinition}}
\newcommand{\ED}{\end{mydefinition}}
\newcommand{\BPF}{\begin{proof}}
\newcommand{\EPF}{\qed \end{proof}}
\newcommand{\BEX}{\begin{myexample}}
\newcommand{\EEX}{\end{myexample}}
\newcommand{\BOB}{\begin{myobservation}}
\newcommand{\EOB}{\end{myobservation}}
\newcommand{\BAL}{\begin{myalgorithm}}
\newcommand{\EAL}{\end{myalgorithm}}

\newcommand{\BE}{\begin{enumerate}}
\newcommand{\EE}{\end{enumerate}}
\newcommand{\BI}{\begin{itemize}}
\newcommand{\EI}{\end{itemize}}

\newenvironment{Rightitem}{%
  \begin{itemize}}{\end{itemize}}
\newenvironment{Leftitem}{%
    \begin{itemize}}{\end{itemize}}
\newenvironment{Iffitem}{%
    \begin{itemize}}{\end{itemize}}
\newenvironment{Iffitemi}{%
    \begin{itemize}}{\end{itemize}}

\newcommand{\BRI}{\begin{Rightitem}\item}  \newcommand{\ERI}{\end{Rightitem}}
\newcommand{\BLI}{\begin{Leftitem}\item}   \newcommand{\ELI}{\end{Leftitem}}
\newcommand{\BIFF}{\begin{Iffitem}\item}   \newcommand{\EIFF}{\end{Iffitem}}
\newcommand{\BIFFI}{\begin{Iffitemi}\item}   \newcommand{\EIFFI}{\end{Iffitemi}}

\newcommand{\bc}{\begin{center}}
\newcommand {\ec}{\end{center}}

\newcommand{\comment}[1]{}

\renewenvironment{thebibliography}[1]{%
  \begin{oldthebibliography}{#1}%
   \vspace{0.2cm}
    \setlength{\parskip}{0ex}%
    \setlength{\itemsep}{0ex}%
}%
{%
  \end{oldthebibliography}%
}

\title{Optimizing Computation of Recovery Plans\\ for BPEL Applications}

\author{Jocelyn Simmonds \qquad\qquad Shoham Ben-David \qquad\qquad Marsha Chechik
\institute{University of Toronto\\
Toronto, ON, Canada}
\email{\{jsimmond,shoham,chechik\}@cs.toronto.edu}
}
\def\titlerunning{Monitoring and Recovery of BPEL Applications}
\def\authorrunning{J. Simmonds, S. Ben-David \& M. Chechik}

\begin{document}
\maketitle

\begin{abstract}
Web service applications are distributed processes that are composed of dynamically bounded services.   In our previous work~\cite{simmonds10a}, we have described a framework for performing runtime monitoring of web service against behavioural correctness properties (described using property patterns and converted into finite state automata).  These specify forbidden behavior (safety properties) and desired behavior (bounded liveness properties).  Finite execution traces of web services described in BPEL are checked for conformance at runtime.
When violations are discovered, our framework automatically proposes and ranks recovery plans which users can then select for execution.  Such plans for safety violations essentially involve ``going back" -- compensating the executed actions until an alternative behaviour of the application is possible.  For bounded liveness violations,
recovery plans include both ``going back'' and ``re-planning'' -- guiding the application towards a desired behaviour. Our experience, reported in \cite{simmonds10c}, identified a drawback in this approach:  we compute too many plans due to (a) overapproximating the number of program points where an alternative behaviour is possible and (b) generating recovery plans for bounded liveness properties which can potentially violate safety properties.   In this paper, we describe improvements to our framework that remedy these problems and describe their effectiveness on a case study.\vspace{-0.2cm}

\end{abstract}

\section{Introduction}
\label{sec:intro}

A BPEL application is an orchestration of (possibly third-party) web services. These services, which can be written in a variety of languages, communicate through published interfaces. Third-party services can be dynamically discovered, and may be modified without notice. BPEL includes mechanisms for  dealing with termination and for specifying compensation actions (these are defined on a ``per action'' basis, i.e., a compensation for booking a flight is to cancel the booking); yet, they are of limited use since it is hard to determine the state of the application  after executing a set of compensations. 

In~\cite{simmonds10a}, we proposed a framework for runtime monitoring and recovery that uses user-specified behavioral properties to automatically compute recovery plans. This framework takes as input the target BPEL application, enriched with the compensation mechanism that allows us to undo some of the actions of the program, and a set of properties (specified as  desired/forbidden behaviors). When a  violation of a property is detected at runtime, this framework outputs a set of ranked recovery plans and enables applying the chosen plan to continue the execution. Such plans for safety violations consist just of the ``going back'' part, until an alternative behavior of the application is possible. For bounded liveness violations, recovery plans include both the ``going back'' and the ``re-execution'' part -- guiding the application towards a desired behavior (such plans are schematically shown using a dashed line in Figure~\ref{fig:unredo}).

For example, consider the Travel Booking System (\tbs) shown in Figure~\ref{fig:tbs_bpel}, which provides travel booking services over the web. In a typical scenario, a customer enters the expected travel dates, the destination city and the rental car location -- airport or hotel.  The system searches for  available flights, hotel rooms and rental cars, placing holds on the resources that best satisfy the customer preferences. If the customer chooses to rent a car at the hotel, the system also books the shuttle between the airport and the hotel.  If the customer likes the itinerary presented to him/her, the holds are turned into bookings;  otherwise, the holds are released. Some  correctness properties of \tbs are $P_1$: ``there should not be a mismatch between flight and hotel dates'' (expressing a safety property, or a forbidden behavior), $P_2$: ``a car reservation request will be fulfilled regardless of the location (i.e., airport or hotel) chosen'' (expressing a bounded liveness property  or a desired behavior), and  $P_3$: ``ground transportation must not be picked before a flight is reserved" (forbidden behavior).

If the application exhibits a forbidden behavior, our framework suggests plans that use compensation actions to allow the application to ``go back'' to an earlier state at which an alternative path that potentially avoids the fault is available. We call such states ``change states''; these include user choices and  non-idempotent partner calls (i.e., those where a repeated execution with the same arguments may yield a different outcome)~\cite{simmonds10a}. For example, if the \tbs system  produces an itinerary with inconsistent dates, a potential recovery plan might be to cancel the current hotel booking and make a new reservation that is consistent with the booked flight's dates. 

Another possibility is that the system fails to produce a desired behavior when calls to some partners terminate, leaving it in an unstable state.  In such cases, our framework computes plans that redirect the application towards executing new activities, those that lead to completion of the desired behavior. For example, if the car reservation partner for the hotel location fails (and thus the ``shuttle/car at hotel'' combination is not available), the recovery plans would be to provide transportation to the user's destination (her ``goal'' state) either by trying to book another car at the hotel, or by undoing the shuttle reservation and try to reserve the rental car from the airport instead.

Effectiveness and scalability of a recovery framework like ours is in (quickly) generating a small number of highly relevant plans.  While our framework can generate recovery plans as discussed above, in our experience with \tbs, reported in \cite{simmonds10c}, we observed that it generates too many plans.  At least two factors contribute to this problem: 
\begin{enumerate}
\item we over-approximate the set of change states and thus offer plans where compensation cannot produce an alternative path through the original system to avoid an error; and
\item some recovery plans for desired behavior violations will (necessarily) lead to violations of forbidden behaviors when executed, and thus should not be offered to the user. 
\end{enumerate}
In this paper, we present two improvements that try to address these issues. The first improvement identifies the non-idempotent service calls that are relevant to the violation, i.e., their execution may affect the control flow of the current execution. The second improvement identifies computed plans that always lead to violations of forbidden behaviors, as the execution of these plans will cause another runtime violation and thus they should not be offered to the user.

In what follows, we give a brief overview of the framework (Sec.~\ref{sec:overview}) and our previous experience with the Travel Booking System (\tbs) (Sec.~\ref{sec:tbs}). In Sec.~\ref{sec:ext}, we use the \tbs example to discuss the two plan generation improvements as well as their effectiveness.  We conclude in Sec.~\ref{sec:conclusion} with a summary of the paper, related work and suggestions for future work.

\vspace{-0.2cm}
\section{Overview of the Approach}
\vspace{-0.2cm}
\label{sec:overview}

\begin{figure}[!t]
\centering
\subfloat[]{\includegraphics[scale=0.45]{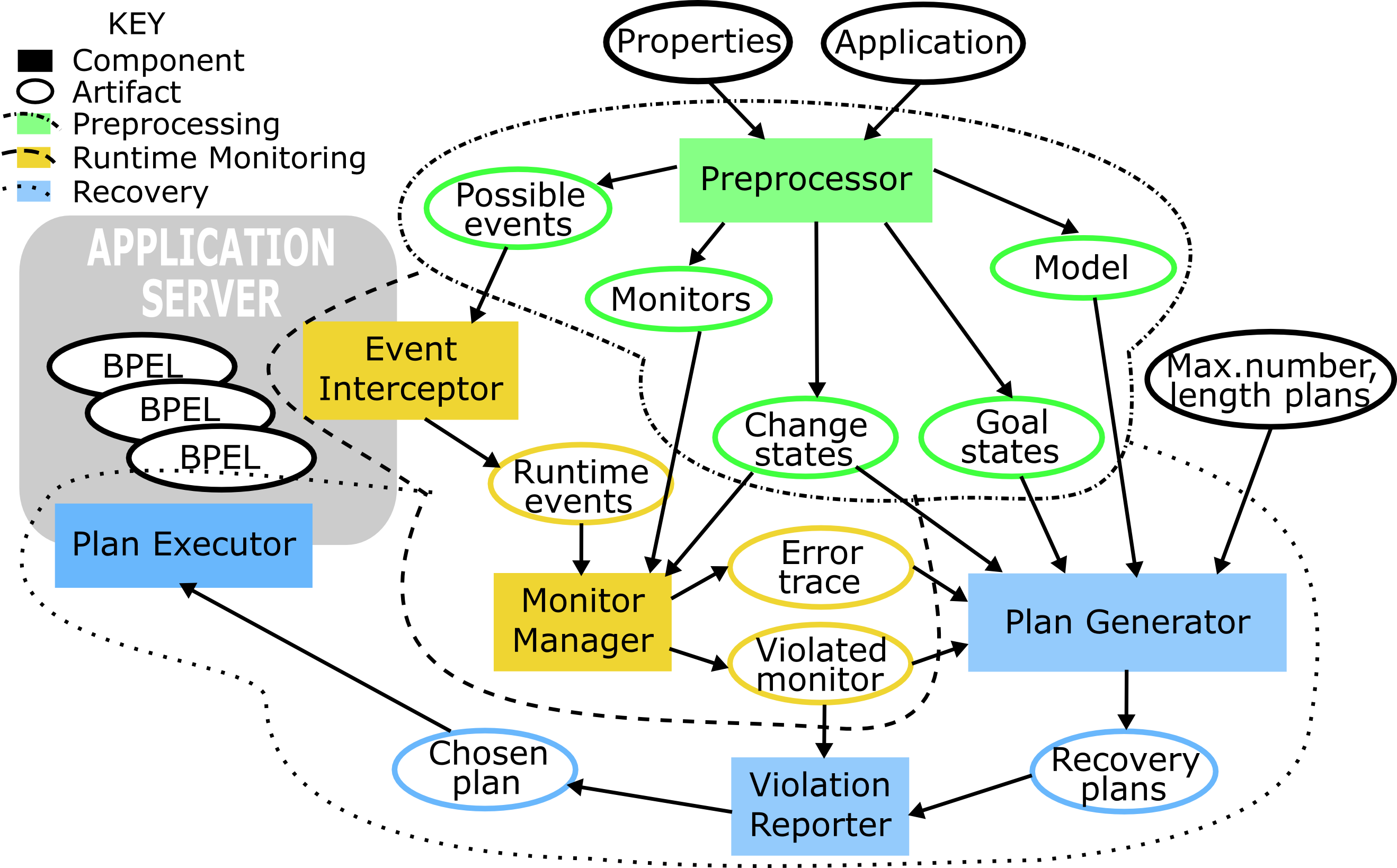}\label{fig:arch}}
\hfil
\subfloat[]{\includegraphics[scale=0.6]{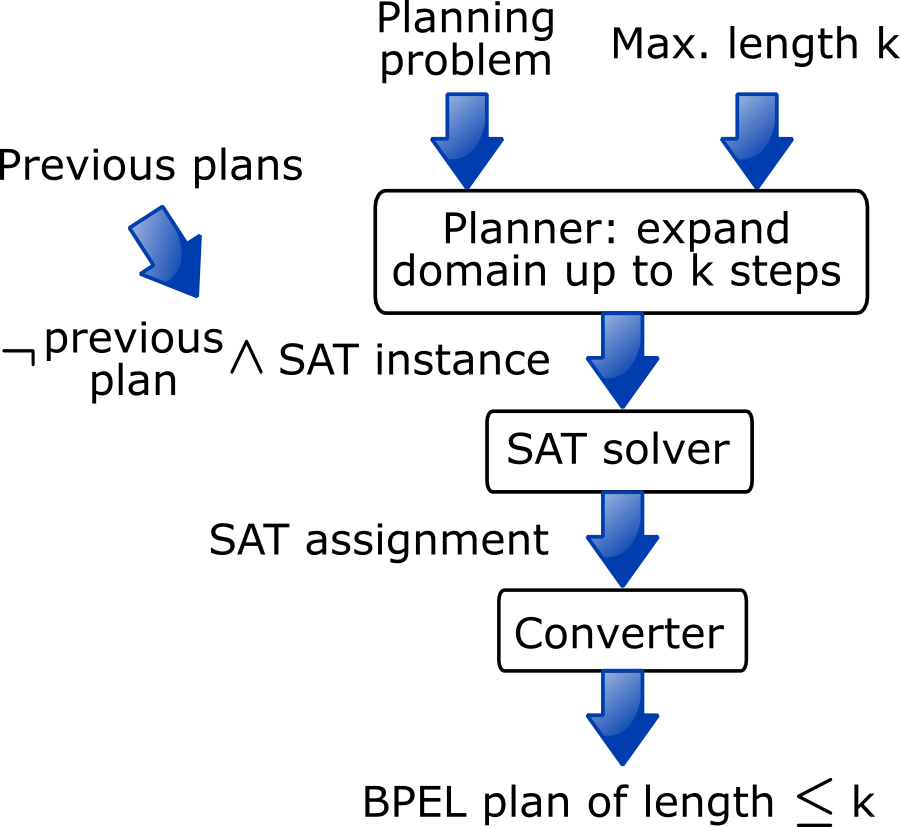}\label{fig:planning}}
\vspace{-0.1cm}
\caption{a) Architecture of the tool; b) Recovery plan generation for violating a desired behavior.}
\vspace{-0.3cm}
\end{figure}

We have implemented our \textsc{RU}ntime \textsc{MO}nitoring and \textsc{R}ecovery framework (\rumor) using a series of publicly available tools and several short (200-300 lines) new Python or Java scripts. The architecture of our tool is shown in Fig.~\ref{fig:arch}, where components and artifacts have been grouped by phase (Preprocessing, Monitoring or Recovery). In the Preprocessing phase, the correctness properties specifying desired and forbidden behaviors are turned into finite-state automata (monitors). We use the WS-Engineer extension for LTSA~\cite{foster06a} to translate the BPEL application into a Labeled Transition System (LTS), enriched with compensation actions (model).  We also compute change and goal states during this phase.

The Monitoring phase is implemented on top of the IBM WebSphere Process Server\footnote{\url{http://www-306.ibm.com/software/integration/wps/}}, a BPEL-compliant process engine for executing BPEL processes. Monitoring is done in a non-intrusive manner -- the Event Interceptor component intercepts runtime events and sends them to the Monitor Manager, which updates the state of the monitors. The use of high-level properties allows us to detect the violation, and our event interception mechanism allows us to stop the application \emph{right before} the violation occurs. \rumor does not require any code instrumentation, does not significantly affect the performance of the monitored system (see \cite{simmonds09}), and enables reasoning about partners expressed in different languages.

During the Recovery phase, the Plan Generator component generates recovery plans using SAT-based planning techniques (see~\cite{simmonds10a} for details). In the case of forbidden behavior violations, the Plan Generator determines which visited change states are reachable by executing available compensation actions. Multiple change states can be encountered along the way, thus leading to the computation of multiple plans. In the case of desired behavior violations, the Plan Generator tries to solve the following planning problem: ``From the current state in the system, find all plans (up to length $k$) to achieve the goal, that go through a change state''. The actions that a plan can execute are defined by the application itself; thus, the domain of the planning problem is the LTS model of the application. The initial and goal states of the planning problem are the current error state and the precomputed goal states, respectively. 

The process for computing recovery plans for desired behavior violations is shown in Fig.~\ref{fig:planning}. \rumor uses Blackbox~\cite{kautzS99}, a SAT-based planner, to convert the planning problem into a SAT instance. The maximum plan length is used to limit how much of the application model is unrolled in the SAT instance, effectively limiting the size of the plans that can be produced. Multiple plans are generated by modifying the initial SAT instance: new plans are obtained by ruling out those computed previously. Plans are extracted from the satisfying assignments produced by the SAT solver SAT4J and converted into BPEL for displaying and execution. SAT4J is an \emph{incremental} SAT solver, i.e., it saves results from one search and uses them for the next.  We take advantage of this for generating multiple plans.

All computed plans are presented to the application user through the Violation Reporter  component.  It generates a web page snippet with violation information as well as a web page for selecting a recovery plan.  The application developer must include this snippet in the default error page, so that the computed recovery plans are displayed as part of the application when an error is detected. The Plan Executor executes the selected plan using dynamic workflows~\cite{vanderaalst05}.  \rumor takes advantage of their implementation as part of IBM WebSphere.

\vspace{-0.2cm}
\section{Monitoring The Travel Booking System}
\vspace{-0.2cm}
\label{sec:tbs}

\subsection{BPEL Model}
\begin{figure}[!t]
\centering
\subfloat{\hspace{-0.3cm}\includegraphics[scale=0.38]{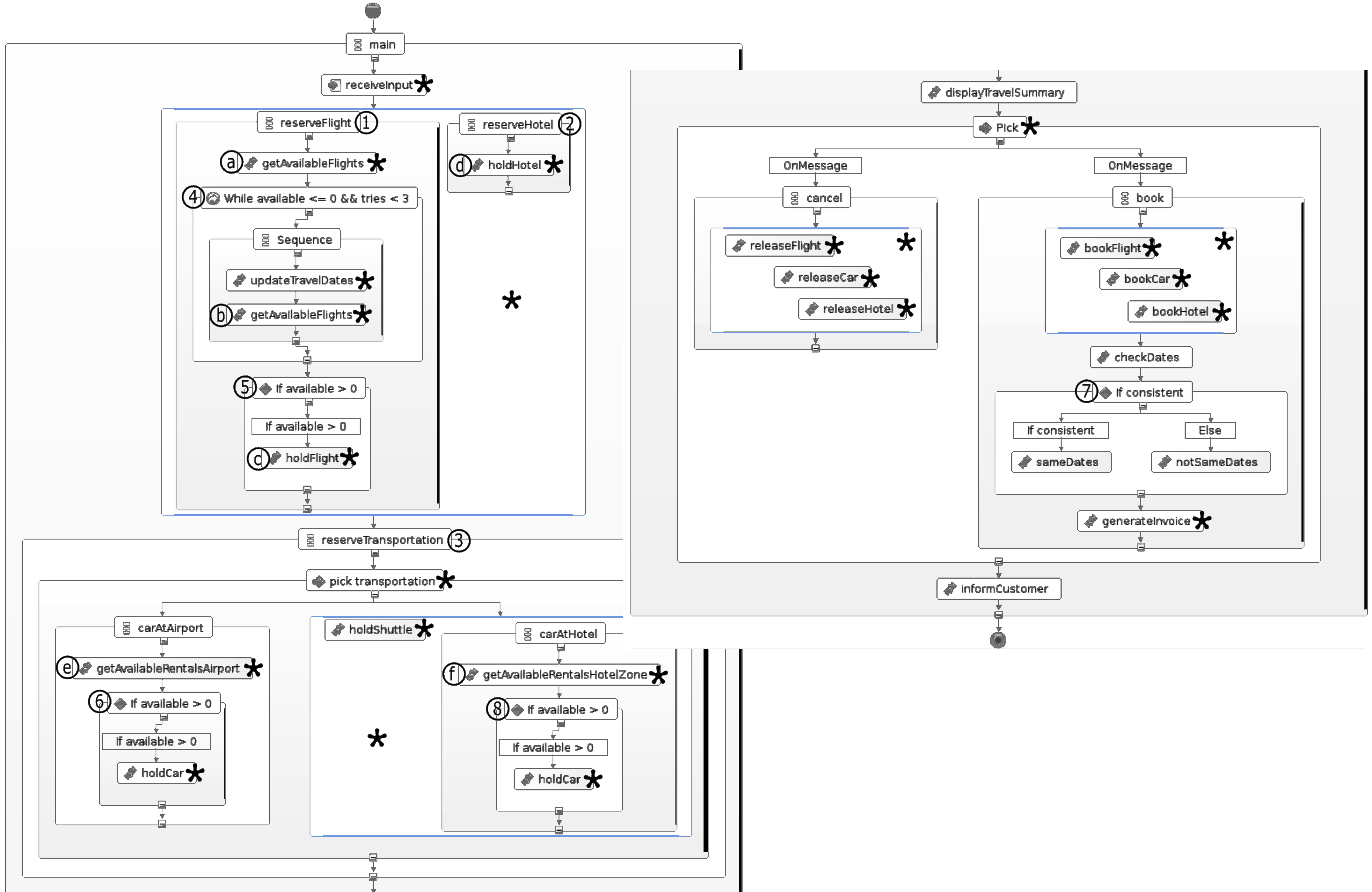}}
\vspace{-0.1cm}
\caption{BPEL implementation of the Travel Booking System. }
\label{fig:tbs_bpel}
\vspace{-0.3cm}
\end{figure}

Figure~\ref{fig:tbs_bpel} shows the BPEL implementation of this system. \tbs interacts with three partners ($\mathsf{FlightSystem}$, $\mathsf{HotelSystem}$ and $\mathsf{CarSystem}$), each offering the services to find an available resource (flight, hotel room, car and shuttle), place a hold on it, release a hold on it, book it and cancel it. Booking a resource is compensated by canceling it and placing a hold is compensated by a release. All other activities can be simply undone, i.e., they do not have explicit compensation actions.  All external service calls are non-idempotent. In the rest of this paper, $\mathsf{bf}$, $\mathsf{bh}$ and $\mathsf{hc}$ represent the service calls $\mathsf{bookFlight}$, $\mathsf{bookHotel}$ and $\mathsf{holdCar}$, respectively. 

The workflow begins by $<$receive$>$'ing input ($\mathsf{receiveInput}$), followed by $<$flow$>$ (i.e., parallel composition) with two branches, since the flight and hotel reservations can be made independently. The branches are labeled \includegraphics{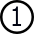} and \includegraphics{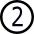}: \includegraphics{./figures/circ1.png}) find and place a hold on a flight, \includegraphics{./figures/circ2.png}) place a hold on a hotel room (this branch has been simplified in this case study). If there are no flights available on the given dates, the system will prompt the user for new dates and then search again (up to three tries). After making the hotel and flight reservations, the system tries to arrange transportation (see the $<$pick$>$ (i.e., making the external choice) activity labeled \includegraphics{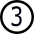}): the user $<$pick$>$'s a rental location ($\mathsf{pickAirport}$ or $\mathsf{pickHotel}$, abbreviated as $\mathsf{pa}$ and $\mathsf{ph}$, respectively) and the system tries to place holds on the required resources (car at airport, or car at hotel and a shuttle between the airport and hotel). 

Once ground transportation has been arranged, the reserved itinerary is displayed to the user ($\mathsf{display-}$ $\mathsf{TravelSummary}$), and at this point, the user must $<$pick$>$ to either $\mathsf{book}$ or $\mathsf{cancel}$ the itinerary. The $\mathsf{book}$ option has a $<$flow$>$ activity that invokes the booking services in parallel, and then calls two local services: one that checks that the hotel and flight dates are consistent ($\mathsf{checkDates}$), and another that generates an invoice ($\mathsf{generateInvoice}$). The result of $\mathsf{checkDates}$ is then passed to local services to determine whether the dates are the same ($\mathsf{sameDates}$) or not ($\mathsf{notSameDates}$, abbreviated as $\mathsf{nsd}$). The $\mathsf{cancel}$ option is just a $<$flow$>$ activity that invokes the corresponding release services. Whichever option is picked by the user, the system finally invokes another local service to inform the user about the outcome of the travel request ($\mathsf{informCustomer}$).

\vspace{-0.2cm}
\subsection{Monitoring Behavioral Properties}

In general, the framework described in \cite{simmonds10a} allows the system
developer to express desired and forbidden behavior as bounded liveness
and safety properties, respectively.  These are expressed using property
patterns~\cite{dwyer98a},
converted into quantified regular expressions (QRE)~\cite{oleander90}
and then become monitoring automata.  
\begin{figure}[!t]
\vspace{-0.15cm}
\centering
\subfloat[]{\includegraphics[scale=1.2]{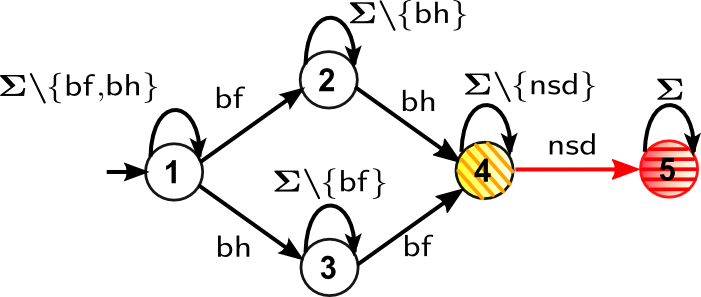}\label{fig:prop1}}
\hfil
\subfloat[]{\includegraphics[scale=1.2]{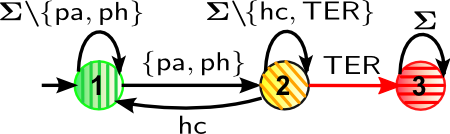}\label{fig:prop2}}
\hfil
\subfloat[]{\includegraphics[scale=1.2]{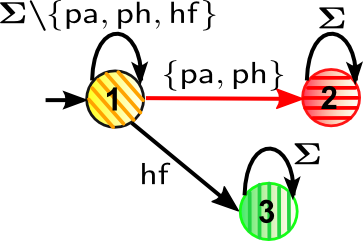}\label{fig:prop3}}
\vspace{-0.2cm}
\caption{Monitors: (a) $\mathtt{A_1}$, (b) $\mathtt{A_2}$ and (c) $\mathtt{A_3}$. Red states are shaded horizontally, green states are shaded vertically, and yellow states are shaded diagonally.}
\vspace{-0.3cm}
\label{fig:props}
\end{figure}
For example,  the \tbs properties described in Sec.~\ref{sec:intro}
are expressed as follows:

\begin{description}
\item [$P_1$:] \textbf{Absence} of a date mismatch event ($\mathsf{notSameDate}$) \textbf{After} both a flight and hotel have been booked ($\mathsf{bookFlight}$ and $\mathsf{bookHotel}$, in any order).\vspace{-0.15cm}
\item [$P_2$:] {\bf Globally} hold a car ($\mathsf{holdCar}$) in \textbf{Response} to a rental location selection ($\mathsf{pickHotel}$ or $\mathsf{pickAirport}$).\vspace{-0.15cm}
\item [$P_3$:] \textbf{Existence} of flight reservation ($\mathsf{holdFlight}$) \textbf{Before} the rental location selection ($\mathsf{pickHotel}$ or $\mathsf{pickAirport}$).
\end{description}

In our framework, monitors are finite state automata that accept \emph{bad} computations. In order to facilitate recovery, we assign colors to the monitor states: 
\begin{itemize}
\item Accepting states are colored red, signaling violation of the property. State 5 of Fig.~\ref{fig:prop1}, state 3 in Fig.~\ref{fig:prop2}, and state 2 in Fig.~\ref{fig:prop3} are red states. \vspace{-0.15cm}
\item Yellow states are those from which a red state can be reached through a single transition. State 4 in Fig.~\ref{fig:prop1}, state 2 in Fig.~\ref{fig:prop2}, and state 1 in Fig.~\ref{fig:prop3} are yellow states. \vspace{-0.15cm}
\item Green states are states that can serve as good places to which a recovery plan can be directed. We define green states to be those states that are not red or yellow, but that can be reached through a single transition from a yellow state. State 1 in Fig.~\ref{fig:prop2}, and state 3 in Fig.~\ref{fig:prop3} are green states.
\end{itemize}\vspace{-0.15cm}
The details of the QRE translation and the formal definition of state colors can be found in~\cite{simmonds10c}. 

The monitor $\mathtt{A_1}$ in Fig.~\ref{fig:prop1} enters its error state ($5$) when the application determines that the hotel and flight booking dates do not match (the hotel and flight can be booked in any order). The monitor $\mathtt{A_2}$ in Fig.~\ref{fig:prop2} represents property $P_2$: if the application terminates (i.e., sends the $\mathsf{TER}$ event) before $\mathsf{hc}$ appears, the monitor moves to the (error) state 3. State 1 is a good state since the monitor enters it once a car has been placed on hold ($\mathsf{hc}$). The monitor $\mathtt{A_3}$ in Fig.~\ref{fig:prop3} represents property $P_3$: it enters the good state 3 once a hold is placed on a flight ($\mathsf{hf}$), and enters its error state 2 if the rental location ($\mathsf{pa}$ or $\mathsf{ph}$) is picked before a flight is reserved ($\mathsf{hf}$). 

\vspace{-0.15cm}
\subsection{From BPEL to LTS}
\vspace{-0.15cm}

In order to reason about BPEL applications, we need to represent them formally, so as to make precise the meaning of ``taking a transition'', ``reading in an event'', etc. Several formalisms for representing BPEL models have been suggested~\cite{fu04,hinz05,ouyang07}. In this work, we use Foster's~\cite{foster06} approach of using a Labeled Transition System (LTS) as the underlying formalism. 

\BD[Labeled Transition Systems] A \emph{Labeled Transition System} LTS is a quadruple $(S, \Sigma, \delta , I) $, where $S$ is a set of states, $\Sigma$ is a set of labels, $\delta \subseteq S \times \Sigma \times   S$ is a transition relation, and $I \subseteq  S$ is a set of initial states.
\ED
Effectively, LTSs are state machine models,  where transitions are labeled whereas states are not. We often use the notation  $s\stackrel{a}{\longrightarrow} s^\prime$ to stand for $(s,a,s^\prime)\in \delta$. An \emph{execution}, or a \emph{trace}, of an LTS is a sequence ${\mathsf T} = s_0 a_0 s_ 1 a_1 s_2... a_{n-1} s_n$  such that $\forall i, 0\leq i < n$, $s_i \in S$, $a_i \in \Sigma$ and  $s_i\stackrel{a_i}{\longrightarrow} s_{i+1}$.

The set of labels $\Sigma$ is derived from the possible application events: service invocations and returns, messages, scope entries, and conditional valuations. \cite{foster06} specifies the mapping of all BPEL 1.0 activities into LTS. Conditional activities like $<$if$>$ and $<$while$>$ statements are represented as states with two outgoing transitions, one for each valuation of the condition. $<$pick$>$ is also a conditional activity, but can have one or more outgoing transition, one for each $<$onMessage$>$ branch. $<$sequence$>$ and $<$flow$>$ activities result in the sequential and parallel composition of the enclosed activities, respectively. 

In~\cite{simmonds10a}, we describe how we augmented Foster's translation so that we can model termination, as well as BPEL compensation. According to our translation, the \tbs LTS has 52 states and 67 transitions, and $|\Sigma| = 33$.  20 of the BPEL activities  (highlighted with a \includegraphics[scale=1.2]{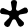} symbol in Figure~\ref{fig:tbs_bpel})  yield a total of 35 change states in the LTS.

\begin{table}[t]
\centering
\scalebox{0.8}{
\begin{tabular}{|c|c|c|c|c|c|c|c|c|c|c|}
\hline
\textbf{Scenario}&  \textbf{k} & \textbf{Change states} &
\textbf{Vars} & \textbf{Clauses} & \textbf{Plans}  & \textbf{Time (s)}  \\
\hline
$t_1$ & 5 & 2 & -- & -- & 2 & 0.01 \\ %
 & 10 & 5 & -- & -- & 5 & 0.02 \\
 & 15 & 8 & -- & -- & 8 & 0.02 \\
 & 20 & 12 & -- & -- & 12 & 0.02 \\
 & 25 & 13 & -- & -- & 13 & 0.02 \\
 & 30 & 13 & -- & -- & 13 & 0.02 \\
\hline
$t_2$ & 5 & 4 & 108 & 464 & 0 & 0.01 \\ %
 & 10 & 7 & 883 & 30,524 & 2 & 0.14 \\
 & 15 & 10 & 1,456 & 74,932 &  8 & 1.37 \\
 & 20 & 10 & 2,141 & 135,047 &  18 & 4.72 \\
 & 25 & 10 & 3,298 & 246,210 &  60 & 29.16 \\
 & 30 & 10 & 5,288 & 464,654 &  68 & 61.34 \\
\hline
\end{tabular}}
\caption{Plan generation data.  ``--'' mark cases which are not applicable, such as references to SAT for recovery from forbidden behavior violations. \label{tab:original_data}}
\vspace{-0.3cm}
\end{table}

\vspace{-0.15cm}
\subsection{Experience: Recovery from a safety property violation}
\label{sec:exp1}
\vspace{-0.15cm}

We generated a recovery plan for the following scenario (called trace $t_1$, of length $k=21$) which violates property $P_1$: The application first makes a hotel reservation ($\mathsf{holdHotel}$) and then prompts the user for new travel dates ($\mathsf{updateTravelDates}$), since there were no flights available on the current travel dates. The car rental location is the airport ($\mathsf{pickAirport}$). The system displays the itinerary ($\mathsf{displayTravelSummary}$) but the user does not notice the date inconsistency and decides to $\mathsf{book}$ it. The $\mathtt{TBS}$ makes the bookings ($\mathsf{bookFlight}$, $\mathsf{bookHotel}$ and $\mathsf{bookCar}$) and then checks for date consistency ($\mathsf{checkDates}$). Since the dates are not the same ($\mathsf{notSameDates}$), we detect violation of $P_1$ and initiate recovery.

We generated plans starting with length $k=5$ and going to $k=30$ in increments of 5. In order to generate all possible plans for each $k$, we chose $n$ -- the maximum number of plans generated -- to be $\mathtt{MAX\_INT}$.  Table~\ref{tab:original_data} summarizes the results. A total of 13 plans were generated, and the longest plan, which reaches the initial state, is of length 21 (and thus the rows corresponding to $k=25$ and $k=30$ contain identical information).
Since $t_1$ violates a safety property, no SAT instances were generated, and the running time of the plan generation is trivial.

The following plans turn $t_1$ into a successful trace: $p^1_A$ -- cancel the flight reservation and pick a new flight using the original travel dates, and $p^1_B$ -- cancel the hotel reservation and pick a new hotel room for the new travel dates. Our tool generated both of these plans, but ranked them 11th and 12th (out of 13), respectively. They were assigned a low rank due to the interplay between the following two characteristics of our case study: (i) the actual error occurs at the beginning of the scenario (in the flight and hotel reservation $<$flow$>$), but the property violation was only detected near the end of the workflow (in the $\mathsf{book}$ flow), and (ii) $t_1$ passes through a relatively large number of change states, and thus many recovery plans are possible.  

The first of these causes could be potentially fixed by writing ``better'' properties -- the ones that allows us to catch an error as soon as it occurs.  We recognize, of course, that this can be difficult to do. The second stems from the fact that not all service calls marked as non-idempotent are relevant to $P_1$ or its violation. In Sec.~\ref{sec:ext_change}, we present a method for identifying the non-idempotent service calls that are relevant to the violation, i.e., their execution may affect the control flow of the current execution. By reducing the number of change states considered, fewer recovery plans will be generated.

\vspace{-0.15cm}
\subsection{Experience: Recovery from a bounded liveness property violation}
\label{sec:exp2}
\vspace{-0.15cm}

The following scenario (we call it trace $t_2$, with length 14) violates property $P_2$.  Consider an execution where the user reserves a hotel room ($\mathsf{reserveHotel}$), and  a flight ($\mathsf{reserveFlight}$). He then  chooses to rent a car at the hotel ($\mathsf{pickHotel}$), but no cars are available at that hotel. $\mathtt{TBS}$ makes flight, hotel and shuttle reservations ($\mathsf{holdFlight}$ and $\mathsf{holdHotel}$), but never makes a car reservation ($\mathsf{holdCar}$). The user does not notice the missing reservation in the displayed itinerary ($\mathsf{displayTravelSummary}$) and decides to $\mathsf{book}$ it. The $\mathtt{TBS}$ tries to complete the bookings, first booking the hotel ($\mathsf{bookHotel}$) and then the car ($\mathsf{bookCar}$). When the application attempts to invoke $\mathsf{bookCar}$, the BPEL engine detects that the application tries to access a non-initialized process variable (since there is no car reservation), and issues a $\mathsf{TER}$ event.  Rather than delivering this event to the application, we initiate recovery.

We are again using $n = \mathtt{MAX\_INT}$ and varying $k$ between 5 and 30, in increments of 5, summarizing the results in  Table~\ref{tab:original_data}.  The first thing to note is that our approach generated a relatively large number of plans (over 60) as $k$ approached 30. While in general the further  we move away from a goal link, the more alternative paths lead back to it, this was especially true for $\mathtt{TBS}$ which  had a number of $<$flow$>$ activities. The second thing to note is that our analysis remained tractable even as the length of the plan and the number of plans generated grew (around 1 min for the most expensive configuration).

Executing one of the following plans would leave $\mathtt{TBS}$ in a desired state: $p^2_A$ -- attempt the car rental at the hotel again, and $p^2_B$ --  cancel the shuttle from the airport to the hotel and attempt to rent a car at the airport. Unlike $t_1$, the error in this scenario was discovered soon after its occurrence, so plans $p^2_A$ and $p^2_B$ are the first ones  generated by our approach. $p^2_A$ actually corresponds to two plans, since the application logic for reserving a car at a hotel is in a $<$flow$>$ activity, enabling two ways of reaching the same goal link.   Plan $p^2_B$ was the 3rd plan generated. 

The rest of the plans we generated compensate various parts of $t_2$, and then try to reach one of the three goal links. While these longer plans include more compensations and are ranked lower than $p^2_A$ and $p^2_B$, we still feel that it may be difficult for the user to have to sift through all of them. As in the case of safety property violations, we can reduce the number of plans generated by picking relevant change states. Furthermore, some of the computed recovery plans, when executed, lead to violations of safety properties, and thus should not be offered to the user. In Sec.~\ref{sec:ext_forb}, we present a method for identifying such recovery plans that always lead to violations of safety properties.

\vspace{-0.15cm}
\section{Reducing the Number of Generated Plans}
\label{sec:ext}

As discussed above, our tool produces a set of recovery plans for each detected violation.   However, in some cases this set includes unusable
plans.  In this section, we look at techniques for filtering out
two types of unusable plans:  those that require going through
unnecessary change states, where
 re-executing the partner
call cannot affect the (negative) outcome of the trace
(see Sec.~\ref{sec:ext_change}), and
plans that fix a liveness property at the expense of violating
some safety properties (see Sec.~\ref{sec:ext_forb}).

\subsection{Relevant Change States}
\label{sec:ext_change}

As discussed before, change states are application states from which \emph{flow-changing} actions can be executed. These are user choices ($<$pick$>$),  modeling the $<$flow$>$ activity, and service calls whose outcomes are not completely determined by their input parameters (to which we refer as non-idempotent). For example, $\mathsf{getAvailableFlights}$ is a non-idempotent service call (and leads to the identification of various change states), since each new invocation of the service, with the same travel dates, may return different available flights. Non-idempotent service calls are identified by the developer.

Let us reexamine the trace $t_1$. This trace visited 13 change states, of which 11 correspond to non-idempotent service calls. The two flow activities executed on the trace identify two change states that coincide with two states already identified using non-idempotent service calls ($\mathsf{holdHotel}$ and $\mathsf{bookCar}$). The remaining two change states correspond to the two $<$pick$>$ activities on the trace (choice between rental locations, and choice between booking/canceling the itinerary).

As $<$pick$>$ and $<$flow$>$ activities are flow-altering actions by definition, the change states identified by these activities are always relevant to the current violation. On the other hand, not all service calls marked as non-idempotent are relevant, i.e., their execution cannot modify the current execution trace. For example, $\mathsf{bookFlight}$ and $\mathsf{bookHotel}$ are both non-idempotent service calls that appear in $t_1$, and so define two recovery plans. However, these two plans are not useful: after their execution, the application is forced to complete the execution of $t_1$ in its entirety. This happens because none of the later control predicates depend on the output produced by these service calls. This example suggests a definition of \emph{relevant} change state:

\BD[Relevant Change State] A change state is relevant if it is identified by:  1)
  a $<$flow$>$ or $<$pick$>$ activity, or 2)
  a non-idempotent service call, and a variable that appears in a control activity is data dependent on
the outcome of this service call.
\ED

In order to carry out the data dependency analysis on the application LTS, we must first determine which BPEL activities define and use process variables, and how to map this information to the LTS model. $<$invoke$>$ and $<$assign$>$ activities both define and use variables. For example, the $\mathsf{getAvailableFlights}$ service call takes as input the $\mathsf{travelRequest}$ variable (use) and modifies the $\mathsf{available-}$ $\mathsf{Flights}$ variable (definition). Both $<$while$>$ and $<$if$>$ activities use the variables that appear in the activity's predicate. $<$flow$>$ and $<$pick$>$ do not use or define variables.

We can now define the following two sets of variables for each LTS transition $(s \stackrel{a}{\longrightarrow} s^\prime)$: the set of variables defined by the action $a$ ($\mbox{Def}(s \stackrel{a}{\longrightarrow} s^\prime)$), and the set of variables used by action $a$ ($\mbox{Use}(s \stackrel{a}{\longrightarrow} s^\prime)$). Formally:

\scalebox{0.9}{
$
\mbox{Def}(s \stackrel{a}{\longrightarrow} s^\prime) =
\left\{
    \begin{array}{ll}
        \{\mathsf{inVar}\} & \mbox{if } a \mbox{ represents \textless invoke ~\dots~ inputVariable = ``inVar'' \dots \textgreater} \\
        \{\mathsf{fromVar}\} & \mbox{if } a \mbox{ represents \textless assign\textgreater\textless from\textgreater fromVar\textless\slash from\textgreater \dots \textgreater} \\
        \emptyset & \mbox{otherwise}
    \end{array}
\right.
$
}
and

\scalebox{0.9}{
$
\mbox{Use}(s \stackrel{a}{\longrightarrow} s^\prime) =
\left\{
    \begin{array}{ll}
        \{\mathsf{outVar}\} & \mbox{if } a \mbox{ represents \textless invoke ~\dots~ outputVariable = ``outVar'' \dots \textgreater} \\
        \{\mathsf{toVar}\} & \mbox{if } a \mbox{ represents \textless assign\textgreater\textless to\textgreater toVar\textless\slash to\textgreater \dots \textgreater} \\
        \{\mathsf{v_1, v_2, \dots v_n}\} & \mbox{if } a \mbox{ represents a \textless while\textgreater ~or \textless if\textgreater ~branch, and } \{\mathsf{v_1, v_2, \dots v_n}\} \\
            & ~~~~~~  \mbox{ appear in the corresponding \textless condition\textgreater} \\
        \emptyset & \mbox{otherwise}
    \end{array}
\right.
$
}

The set of variable definitions that occur on a trace is the union of the definitions that occur on the individual transitions of the trace: for a trace ${\mathsf T} = s_0 a_0 s_1 a_1 s_2 \dots a_{n-1} s_n$, $\mbox{Def}({\mathsf T}) = \bigcup_i \mbox{Def}(s_i\stackrel{a_i}{\longrightarrow} s_{i+1})$. Now we can define \emph{direct data dependency}: a transition $v$ is directly data dependent on another transition $u$ if and only if $v$ uses a variable defined by $u$, and there is a path from $u$ to $v$ where this variable is not redefined. Formally,

\BD[Directly Data Dependent]\label{def:ddd} A transition $(q \stackrel{b}{\longrightarrow} q^\prime)$ is \emph{directly data dependent} on a transition $(p \stackrel{a}{\longrightarrow} p^\prime)$ if and only if there is a trace ${\mathsf T} = s_0 a_0 s_1 a_1 s_2 \dots a_{n-1} s_n$ such that $p^\prime = s_0$, $q = s_n$ and $\big(\mbox{Def}(p \stackrel{a}{\longrightarrow} p^\prime) \bigcap \mbox{Use}(q \stackrel{b}{\longrightarrow} q^\prime)\big) - \mbox{Def}({\mathsf T}) \neq \emptyset$.
\ED

For example, the $<$if$>$ activity labeled \includegraphics[scale=0.5]{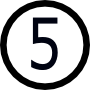} and the $\mathsf{holdFlight}$ service call labeled \includegraphics[scale=0.5]{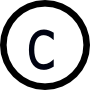} are both directly data dependent on the $\mathsf{getAvailableFlights}$ service calls at \includegraphics[scale=0.5]{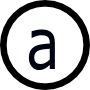} and \includegraphics[scale=0.5]{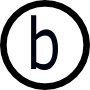}.

We can now  define \emph{data dependency}: a transition $v$ is data dependent on another transition $u$ if there exists a path from $u$ to $v$ that can be divided into sections, where each section is directly data dependent on a previous section. For example, the $\mathsf{bookFlight}$ service call is directly data dependent on the invocation of $\mathsf{holdFlight}$, so $\mathsf{bookFlight}$ is data dependent on both invocations of the $\mathsf{getAvailableFlights}$ service.

\begin{table}[t]
\centering
\scalebox{0.8}{
\begin{tabular}{|c|c|c|c|c|c|}
\hline
\textbf{Trace} & \textbf{Id} & \textbf{Activity}& \textbf{Label}&  \textbf{Predicate} & \textbf{Non-idempotent service calls} \\
\hline
$t_1$ & 1 & $<$while$>$ & \includegraphics[scale=0.5]{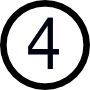} & $\mathsf{availableFlights <= 0 \&\& tries < 3}$ & $\mathsf{getAvailableFlights}$ labeled \includegraphics[scale=0.5]{./figures/circ_a.png}, \includegraphics[scale=0.5]{./figures/circ_b.png} \\
$t_1$ & 2 & $<$if$>$ & \includegraphics[scale=0.5]{./figures/circ5.png} & $\mathsf{availableFlights > 0}$ & $\mathsf{getAvailableFlights}$ labeled \includegraphics[scale=0.5]{./figures/circ_a.png}, \includegraphics[scale=0.5]{./figures/circ_b.png}\\
$t_1$ & 3 & $<$if$>$ & \includegraphics[scale=0.5]{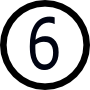} & $\mathsf{availableCars > 0}$ & $\mathsf{getAvailableRentalsAirport}$ labeled \includegraphics[scale=0.5]{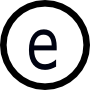}\\
$t_1$ & 4 & $<$if$>$ & \includegraphics[scale=0.5]{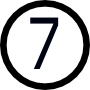} & $\mathsf{consistent == true}$ & $\mathsf{holdHotel, holdFlight}$, labeled \includegraphics[scale=0.5]{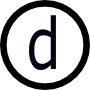}, \includegraphics[scale=0.5]{./figures/circ_c.png}, respectively\\
$t_2$ & 5 & $<$if$>$ & \includegraphics[scale=0.5]{./figures/circ6.png} & $\mathsf{availableFlights <= 0}$ & $\mathsf{getAvailableFlights}$ labeled \includegraphics[scale=0.5]{./figures/circ_a.png}\\
$t_2$ & 6 & $<$if$>$ & \includegraphics[scale=0.5]{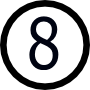} & $\mathsf{availableCars <= 0}$ & $\mathsf{getAvailableRentalsHotel}$ labeled \includegraphics[scale=0.5]{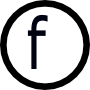}\\
\hline
\end{tabular}}
\caption{Predicates appearing on both traces, and the non-idempotent service calls that affect their values. \label{tab:cond}}
\vspace{-0.6cm}
\end{table}

Now apply data dependency analysis on trace $t_1$. This trace executed four control activities: 1) the $<$while$>$ labeled \includegraphics[scale=0.5]{./figures/circ4.png}, 2) the $<$if$>$ labeled \includegraphics[scale=0.5]{./figures/circ5.png}, 3) the $<$if$>$ labeled \includegraphics[scale=0.5]{./figures/circ6.png}, and 4) the $<$if$>$ labeled \includegraphics[scale=0.5]{./figures/circ7.png}. Table~\ref{tab:cond} lists the corresponding predicates, as well as the non-idempotent service calls that can affect the values of these predicates. For example, the $<$while$>$ condition is $\mathsf{availableFlights <= 0 \&\& tries < 3}$. This use of the $\mathsf{availableFlights}$ variable is directly data dependent on both appearances of the non-idempotent $\mathsf{getAvailableFlights}$ service. On the other hand, the $\mathsf{tries}$ variable is not data dependant on any non-idempotent service calls, since it is  updated by an $<$assign$>$ statement inside the $<$while$>$ activity. 

The data dependency analysis for predicates 2 and 3 is similar to that of predicate 1, and the results of the analysis are summarized in Table~\ref{tab:cond}. In the case of predicate 4, the variable $\mathsf{consistent}$ is directly data dependent on the idempotent service $\mathsf{checkDates}$, which is directly data dependent on the non-idempotent service calls $\mathsf{holdHotel}$ and $\mathsf{holdFlight}$ (since these services modify $\mathsf{reservationData}$, the input parameter of the $\mathsf{checkDates}$ service). 

So, only five of the 10 non-idempotent service calls on trace $t_1$ are identified as relevant. The $<$flow$>$ and $<$pick$>$ activities on trace $t_1$ identify another three relevant change states, so \rumor now generates a total of 0 ($k=5$), 2 ($k=10$), 5 ($k=15$) and 8 ($k=20, 25, 30$) plans for this trace. The desired plans $p^1_A$ and $p^1_B$ are still generated (at $k=20, 25, 30$), but are now ranked 6th and 7th (instead of 11th and 12th). These two plans are still ranked low because of the amount of compensation they require, but by omitting plans that cannot alter the control flow of the current execution, we reduced the number of plans presented to the user by 50\%. 

We carried out the same analysis on trace $t_2$: six of the original 10 change states are marked as relevant. Since trace $t_2$ visits the same $<$pick$>$ and $<$flow$>$ activities as $t_1$, four of the relevant change states are those identified by these activities, the remaining two relevant change states correspond to the non-idempotent service calls associated to predicates 5 and 6, also summarized in Table~\ref{tab:cond}.

\begin{table}[t]
\parbox [t]{0.22\textwidth}{
{\includegraphics[scale=0.6]{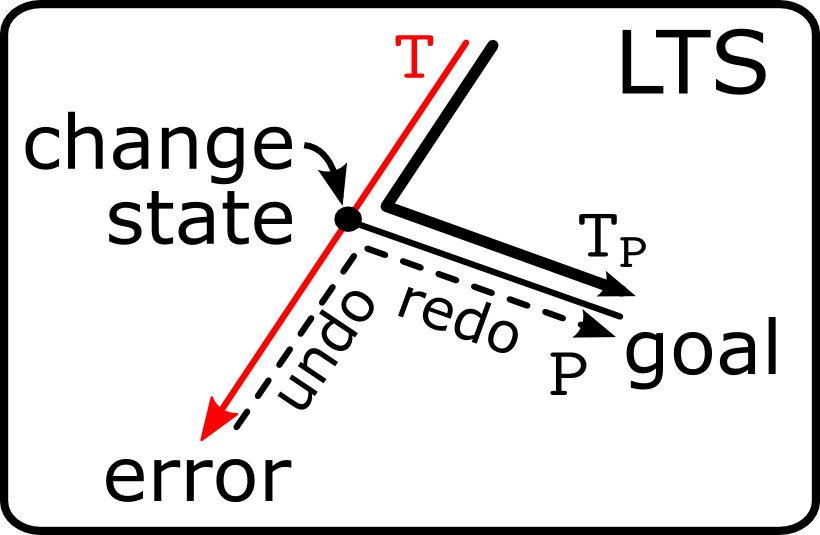}}
\captionof{figure}{A schematic view on plan generation and filtering.}
\label{fig:unredo}
}
\hspace{0.5cm}
\parbox [t]{0.74\textwidth }{
\scalebox{0.78}{
\begin{tabular}{|c|r|r|r|r|r|r|r|}
\hline
\textbf{Scenario}&  \textbf{k} & \multicolumn{2}{c|}{\textbf{Baseline}} &
\multicolumn{2}{c|}{\textbf{Relevant}} & \textbf{Avoiding Forbidden} & \textbf{Both~~~~~~~~~} \\
 & & \multicolumn{2}{c|}{\textbf{(from Table\ref{tab:original_data})}} & \multicolumn{2}{c|}{\textbf{Change States}} & \textbf{Behaviors~~~~~~~~} & \textbf{improvements} \\
\cline{3-8}
 & & \textbf{change} & \textbf{plans} & \textbf{change} & \textbf{plans}& \textbf{plans~~~~~~~~~~~~} & \textbf{plans~~~~~~~} \\
 & & \textbf{states} &  & \textbf{states} & &  &  \\
\hline
$t_1$ & 5  & 2  & 2  & 0 & 0 & --  & -- \\ 
      & 10 & 5  & 5  & 2 & 2 & -- & -- \\
      & 15 & 8  & 8  & 5 & 5 & -- & -- \\
      & 20 & 12 & 12 & 8 & 8 & -- & -- \\
      & 25 & 13 & 13 & 8 & 8 & -- & -- \\
      & 30 & 13 & 13 & 8 & 8 & -- & -- \\
\hline
$t_2$ & 5 & 4  & 0  & 2 & 0 & 0 & 0 \\ %
     & 10 & 7  & 2  & 4 & 2 & 2 & 2 \\
     & 15 & 10 & 8  & 6 & 5 & 8 & 5 \\
     & 20 & 10 & 18 & 6 & 15 & 11 & 8 \\
     & 25 & 10 & 60 & 6 & 41 & 32 & 23 \\
     & 30 & 10 & 68 & 6 & 41 & 38 & 23 \\
\hline
\end{tabular}}
\captionof{table}{Results of applying both improvements (separately, and then combined) to the \tbs case study. ``--'' marks cases which are not applicable, since the second improvement only applies to bounded liveness properties.}
\label{tab:results}
}
\vspace{-0.3cm}
\end{table}

\subsection{Avoiding Forbidden Behaviors}
\label{sec:ext_forb}

Our second method aims to remove those plans that result in the system performing behavior which is explicitly forbidden.  That is, we use safety properties to help filter recovery plans for liveness properties. This process is outlined in Figure~\ref{fig:unredo}:  given a failing trace $\mathtt{T}$, we compute a plan $\mathtt{P}$ which first
``undoes'' the trace until a change state and then computes an alternative path to a certain goal (shown using dashed lines).  $\mathtt{P}$ is \emph{unsuitable} if the path from the initial state going through this change state and continuing via the computed alternative path towards the goal (shown using a thick line and denoted $\mathtt{T_P}$) is forbidden.  That is, there exists a safety monitor $\mathtt{A}_i$ which enters an error state when executed on  $\mathtt{T_P}$.

The simplest method, presented here, applies the filtering w.r.t. safety properties \emph{after} the set of recovery plans has already been produced. That is, given a trace $\mathtt{T}$ and a plan $\mathtt{P}$, we can compute $\mathtt{T_P}$ and simulate every safety monitor on it, removing $\mathtt{P}$ from consideration if any monitor fails.

The path from the initial state to the change state used in $\mathtt{P}$ can be very long, and thus we feel that simulating each monitor on the entire trace $\mathtt{T_P}$ is very inefficient. We also cannot execute monitors backwards from the error state of $\mathtt{T}$ along the ``undo'' part of $\mathtt{P}$: while our monitors are deterministic, their inverse transition relations do not have to be deterministic, making the execution in reverse problematic.

Instead, we aim to maintain enough data during the execution of the trace $\mathsf{T}$ in order to be able to restart monitors directly from the change state, moving along the new, recomputed path of the plan.  To do so, as $\mathtt{T}$ executes, we record the states of all monitors in the system in addition to the states and transitions of the application. Thus, for each state $s$ of the application reached during the execution of trace $\mathtt{T}$, we store  a tuple $(s, s_\mathtt{A_1}, ... , s_\mathtt{A_n})$, where $s_\mathtt{A_i}$ is a state of the monitor $\mathtt{A_i}$ as the application is in state $s$. To check whether $\mathtt{P}$ is a valid plan, we go directly to the change state $s_c$ in $\mathtt{P}$, extract the tuple $(s_c, s_\mathtt{A_1}, s_\mathtt{A_2}, ..., s_\mathtt{A_n})$ stored as part of $\mathtt{T}$ and then simulate each safety monitor $\mathtt{A_i}$ starting it from the state $s_\mathtt{A_i}$ along $\mathtt{P}$ which starts at state $s_c$.

As an example, consider the TBS system and trace $t_2$, described in Sec.~\ref{sec:exp2}, violating the property $P_2$. Our approach produces over 60 plans to recover from this violation, for plan lengths $k \geq 25$ (see Table~\ref{tab:original_data}). Consider the plan that goes back all the way until encountering the change state associated with the call to $\mathsf{getAvailableFlight}$, canceling the booked flights on the way. Afterwards, this plan attempts to re-book a flight, but fails to do so. It continues executing, and tries to pick a car at the airport instead. However, this plan violates property $P_3$ (i.e., monitor $\mathtt{A_3}$ would enter its error state upon seeing an action $\mathsf{pa}$).  Thus, we automatically filter this plan out  and do not present it to the user.

Overall, applying this approach to recovery for trace $t_2$ reduces the number of plans from over 60 to 41. Furthermore, combining it with the computation of the relevant change states, the number of plans is further reduced to 23 (see Table~\ref{tab:results}). While this number is still relatively large, it presents a considerable improvement and enables the user to pick a desired plan more easily.

\vspace{-0.2cm}
\section{Summary and Related Results}
\label{sec:conclusion}

In this paper, we briefly summarized the \rumor approach to runtime monitoring and recovery of web services w.r.t. behavioral properties expressed as desired or forbidden behaviors.  We have also described two optimizations to the recovery plan generation:  reducing the number of change states and using monitors  to filter those plans which represent forbidden behaviors.

Hall{\'e} and Villemaire in~\cite{halle08,halle09b}, suggest a monitoring framework where data-aware properties are written in LTL enriched with first-order quantifications.  Generating automata for runtime monitoring w.r.t. such an expressive language is significantly more complex than in our framework. Recovery in the work of Hall{\'e} and Villemaire is based on executing a predefined function, associated with an individual property -- i.e., all failures of the same property are treated in the same way, statically. In contrast, our method is dynamic and  generates recovery plans customized for each error.

An emerging research area in recent years is that of \emph{self-adaptive} and \emph{self-managed} systems (see \cite{brunM07,KramerM07,chengLGGLMMT09} for a partial list). A system is considered self-adaptive if it is capable of adjusting itself in response to a changing environment.  This approach is different from ours, since our framework does not change  the system itself, and recovery plans are discovered and executed using the original application.

The work of  Carzaniga et al.\cite{carzaniga08} is the closest to ours in spirit. It exploits redundancy in web applications to find workarounds when errors occur, assuming that the application is given as a finite-state machine, with an identified  error state as well as the ``fallback'' state to which the application should return. The approach generates all possible recovery plans, without prioritizing them. In contrast, our framework not only detects runtime errors but also calculates goal and change states and in addition automatically filters out unusable recovery plans. 

Our work in this space is on-going.  Specifically, we are interested in further case studies, optimized usage of SAT solving for better plan generation (e.g., so that we encode forbidden behaviors as part of the SAT problem rather than filtering them out after the plan has been generated), ways to harvest and effectively express behavioral properties, since this is key to the usability of our approach.

\vspace{-0.35cm}

{\small
\bibliographystyle{abbrv}
\bibliography{combined}
}

\end{document}